\documentclass[twocolumn,amssymb,nobibnotes,aps,superscriptaddress]{revtex4-1} 
\usepackage{epsfig,amsmath,amssymb,bm,epsf,graphics,graphicx,verbatim,subfigure,color}
\usepackage{float}
\usepackage{graphicx}
\usepackage{slashed}
\usepackage{graphicx}% Include figure files
\usepackage{dcolumn}% Align table columns on decimal point
\usepackage{bm}% bold math
\usepackage{amsmath}
\usepackage{bbold}
\usepackage[toc,page]{appendix}

\def\rvec{\mathbf{r}} % The original position of a piece of the system
\def\pvec{\mathbf{p}}
\def\bvec{\mathbf{b}}
\def\rvecc{r} % A component thereof
\def\uvec{\mathbf{u}} % The displacement field
\def\uvecc{u} % A component of above
\def\bvec{\mathbf{b}} % The vector from one end of a bond to another
\def\bvecc{b} % A component thereof
\def\spos{\mathbf{p}} % The position of the center of a spring in the unit cell
\def\sposc{p} % A component of above
\def\svec{\mathbf{v}} % The vector along the length of a spring
\def\strain{\epsilon} % the strain tensor
\def\sind{m} % the index of a bond within the cell
\def\ext{e} % a spring extension
\def\rig{\mathbf{R}} % The rigidity map
\def\rigc{R} % a component of above
\def\en{E} % energy
\def\dim{d} % dimension
\def\nvec{\hat{n}} % outward facing surface normal
\def\stress{\sigma} % the stress tensor
\def\eq{\mathbf{Q}} % the equilibrium map
\def\eqc{Q} % a component thereof
\def\evec{\mathbf{e}} % a vector of spring extensions
\def\svec{\mathbf{s}} % a self stress
\def\qvec{\mathbf{q}} % the wavevector
\def\qvecc{q} % a component thereof
\def\qh{\hat{q}} % normalized wavector
\def\fvec{\mathbf{f}} % the force vector
\def\det{\textrm{det}} % the determinant of the rigidity map
\def\ang{\theta_n} % the angle along which topological polarization is calculated

\def\gvec{\mathbf{g}} %the configuration of a lattice, i.e. position of the sites 
\def\lvec{\mathbf{l}} %the lattice vector
\def\bcoef{\beta} % the coefficient beta involved in the decay length
\def\invd{\kappa} %inverse decay rates
\def\dl{\zeta} %decay length

\def \epij {\epsilon_{ij}} %the strain with components ij 
\def \epkl {\epsilon_{kl}} %the strain with components kl
\def \sij {\sigma_{ij}} %the stress with components ij
\def \Rmij {R_{m,ij}} %the mij component of the rigidity map
\def \Qijm {Q_{ij,m}} %the ijm component of the equilibrium map

\def \qh {\hat{q}}  %normalized version of q
\def \rh {\hat{r}} %normalized version of r

\begin{document}

 \title{Topological elasticity of flexible structures}% Force line breaks with \\

 \author{Adrien Saremi}

 \affiliation{School of Physics, Georgia Institute of Technology, Atlanta, Georgia 30332-0430, US}
\author{Zeb Rocklin}
 \affiliation{School of Physics, Georgia Institute of Technology, Atlanta, Georgia 30332-0430, US}

\begin{abstract}
Flexible mechanical metamaterials possess repeating structural motifs that imbue them with novel, exciting properties including programmability, anomalous elastic moduli and nonlinear and robust response. We address such structures via micromorphic continuum elasticity, which allows highly nonuniform deformations (missed in conventional elasticity) within unit cells that nevertheless vary smoothly between cells. We show that the bulk microstructure gives rise to boundary elastic terms. Discrete lattice theories have shown that critically coordinated structures possess a topological invariant which determines the placement of low-energy modes on edges of such a system. We show that in continuum systems a new topological invariant emerges which relates the difference in the number of such modes between two opposing edges. Guided by the continuum limit of the lattice structures, we identify macroscopic experimental observables for these topological properties that may be observed independently on a new length scale above that of the microstructure.
\end{abstract}

\maketitle

\section{Introduction}

Mechanical metamaterials, defined as structures such as origami sheets or spring networks with repeating patterns of elements, possess properties not found in uniform slabs of material, such as superior strength to mass ratios and vanishing (pentamode)~\cite{milton1995elasticity,kadic2012practicability,goodrich2015principle} or even negative (auxetic)~\cite{grima2000auxetic,yang2004review,alderson2007auxetic,HanifpourZap2018} elastic moduli or Poisson ratios. Of particular interest are flexible mechanical metamaterials, which possess low-energy deformation modes that can be used to achieve shape-changing, programmed response and strong nonlinearities~\cite{bertoldi2017flexible}. Some of these properties are now being demonstrated at microscopic length scales, via kirigami (cut) graphene ribbons~\cite{blees2015graphene}, self-assembled patchy colloids~\cite{chen2011directed}, nanolithography~\cite{cha2018experimental} and DNA ``origami''~\cite{rothemund2006folding,castro2011primer}. These cutting-edge techniques raise the question of what happens when such systems are manipulated on scales much larger than the unit cell. This is precisely the limit of conventional solid mechanics, which occurs well above atomic scales. However, conventional elastic theory assumes a smoothly varying strain field, and so cannot capture the short-distance rearrangements of the flexible unit cell.

To address these issues we develop a micromorphic continuum elasticity applicable to flexible structures. While conventional Cauchy elasticity depends only on the strain of an infinitesimal region, micromorphic elasticity considers regions with additional relevant structure~\cite{eringen1968mechanics}. While this is unnecessary for conventional atomic solids, it can lead to much richer mechanical response, somewhat as liquid-crystalline order modifies conventional fluid dynamics. We consider two situations in which such a theory arises: as the long-wavelength limit of microscopic lattice theories, and as an intrinsically continuum theory based solely on macroscopic observables without recourse to a particular microstructure. In either case, the repeating spatial structure gives rise to an energetic term based on the gradient of the elastic strain, which can be integrated to yield a surface term in the elastic energy.

Our particular focus is on \emph{topologically protected boundary modes}. In the lattice theory, it has been shown that a topological invariant derived from the bulk structure controls the number of zero-energy modes on a given surface of mechanically critical lattices (having equal numbers of degrees of freedom and constraints, also called ``Maxwell'' or ``isostatic'') is determined by an integer-valued topological invariant derived from the bulk structure~\cite{kane2014topological,mao2018maxwell} and leading to a wide variety of physical implications~\cite{paulose2015topological,paulose2015selective,chen2016topological,rocklin2017directional,zhang2018fracturing}. This is derived from an integral around the topologically nontrivial Brillouin Zone, which is not present in the continuum limit. Instead, we present analysis resulting in a new invariant, which instead measures the \emph{difference} between the numbers of zero modes on two opposing edges, analogous to the system's polarization rather than its surface charges. Because these systems are only marginally mechanically stable, it becomes important to consider how mechanical constraints generate energy costs based not only on strain but on strain gradients, and these latter terms break spatial inversion symmetry, necessary for the type of polarization we consider.

The remainder of the paper is organized as follows.
In Sec.~\ref{sec:microstructure}, we show how a model class of microstructures generates mechanical constraints that couple both  to macroscopic strains and to microscopic degrees of freedom. 
In Sec.~\ref{sec:surface} we show how this generates surface surface terms complementing bulk ones. In Sec.~\ref{sec:relaxation} we show how a simple linear-algebraic procedure based on the self stresses (stressed states of equilibrium) captures the equilibration of the system and generates an effective theory in terms of smooth strain fields only. In Sec.~\ref{sec:topology} we introduce the topological invariant and relate it to the surface modes, relying on the continuum theory rather than a particular microstructure. In Sec.~\ref{sec:length} we characterize the length and energy scales associated with the boundary modes and how they can be experimentally observed. We conclude in Sec.~\ref{sec:conclusions}.

\section{Effects of smooth strain field on microstructure} \label{sec:microstructure}

\begin{figure}[h]
\center
\includegraphics[scale=0.5]{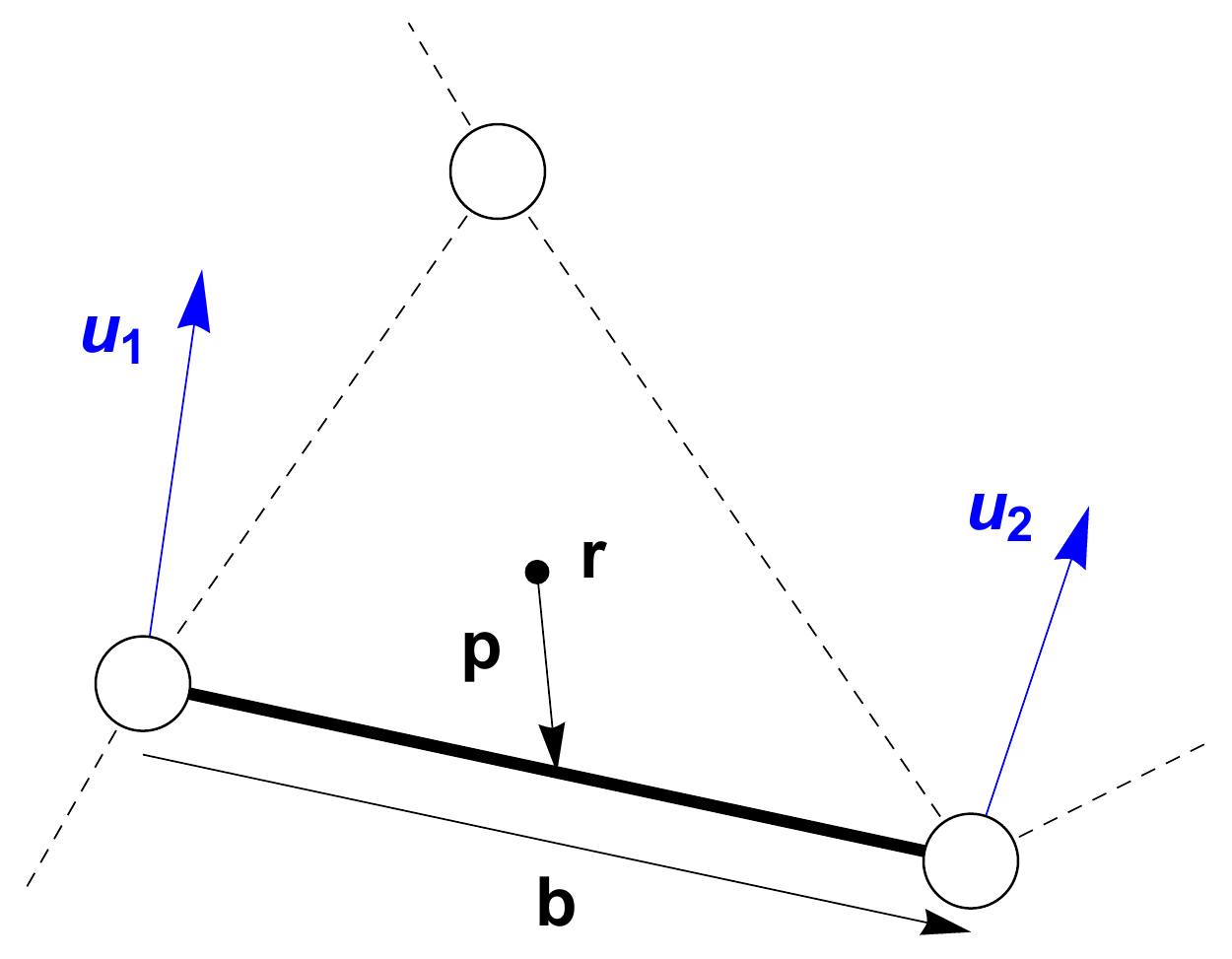}
\caption{
A periodic spring network has a periodic microstructure consisting of bonds connecting sites, as in the generalized kagome lattice shown here. A solid bond is located at $\pvec$ relative to the center $\rvec$ of the cell in which it lies. The bond connects two sites with relative position $\bvec$, undergoing displacements $\uvec_1(\rvec),\uvec_2(\rvec)$ that cause extension/compression of the bond. For continuum fields, the displacements may differ greatly for each site within the repeating cell, but vary smoothly across the cells.
}
\label{fig:diagram}
\end{figure}

Here, we introduce the class of mechanical systems whose energy is stored in discrete, spring-like bonds and consider how these couple to external strain fields. Let $\uvec(\rvec)$ denote the displacements that sites undergo at position $\rvec$ in the undeformed reference space. Consider a particular spring in a periodic crystal cell structure, and let $\rvec$ be the position of its cell, $\pvec$ the position of the center of the bond relative to the cell and $\bvec$ the vector from one end of the bond to the other, as shown in Fig.~\ref{fig:diagram}. The extension of the spring is, for sufficiently smooth displacement fields (repeated \emph{lower} indices implying summation):

\begin{subequations}
\begin{align}
\label{eq:ext}
e &= \frac{1}{|\bvec|} \bvec \cdot \left[ \uvec(\rvec + \spos + \bvec/2)-\uvec(\rvec + \spos - \bvec/2) \right] \\
&\approx \frac{\bvecc_i \bvecc_j}{|\bvec|} \left[ 1+\sposc_k \partial_k  \right] \partial_i \uvecc_j \\
&= \frac{\bvecc_i \bvecc_j}{|\bvec|} \left[ 1+\sposc_k \partial_k \right]  \strain_{ij}.
\end{align}
\end{subequations}

\noindent In the preceding lines, we conduct an expansion assuming that the relative displacements of the ends of the bond remain small, yet considering a higher gradient in the displacement field. Because of the rotational invariance of the problem (reflected in the symmetric $\bvecc_i \bvecc_j$ prefactor) even non-uniform rotations do not extend the spring and instead we obtain a result purely in terms of the symmetrized strain $\strain_{ij} \equiv (1/2)(\partial_i \uvecc_j + \partial_j \uvecc_i)$ and its gradients.  The strain gradient terms are usually ignored in the elasticity of rigid bodies, for which they are small for smooth strain fields. However, it is important in systems near the isostatic point, for which the contributions to the elastic energy of the conventional terms can vanish. Note also that this expression is evaluated at the position of the cell, not the bond, such that our object of interest is the strain field within the cell and the characteristics of the individual bond are encoded in $\spos, \bvec$. 

Notably, given that bonds repeat periodically throughout the structure, there is an ambiguity in which cell to assign which bond, the so-called discrete gauge symmetry of the lattice theory~\cite{kane2014topological}. Choosing a different such assignment would shift $\spos$ by some combination of lattice primitive vectors but shift the point at which the strain is evaluated by an equal and opposite amount, such that the physical observable of the bond extension at a particular point in space (as opposed to a particular cell) is unchanged. Keeping this in mind, we may use the above relationship to construct a tensor that relates the extensions $\ext_\sind(\rvec)$ indexed by $\sind$ to the strain field:

\begin{align}
\label{eq:em}
\ext_\sind(\rvec) &= \rigc^0_{\sind,ij} \strain_{ij}(\rvec) \\
\label{eq:rig}
\rigc^0_{m,ij} &\equiv \frac{\bvecc_i^m \bvecc_j^m}{|\bvec^m|} \left[ 1+\sposc_k^m \partial_k  \right]
\end{align}

\noindent Here, we introduce the \emph{initial rigidity map}, $\rig^0$, which determines the spring extensions resulting from a given set of strains. Although it is the continuum analog of the rigidity (or compatibility) matrix used in lattice theories \cite{lubensky2015phonons, kane2014topological,calladine1978buckminster}, we term it ``initial'' here because it generally results in unbalanced forces on the sites connected via the bonds, a limitation that we resolve below.
Note that since this object maps from dimensionless strains to spring extensions, its elements have units of length. An alternate approach would be to calculate the fractional extensions of the springs, such that the elements of the map were rescaled by spring lengths and thus mapped from the strain of the embedding space to the strains of the springs.

\section{Bulk and surface energies}

\label{sec:surface}

In the preceding Section, we described a purely structural, geometrical relationship between strain fields and mechanical constraints. Here, we relate the violations of these constraints (in particular, the stretching of springs) to elastic energy.
For simplicity, we consider a system with bonds of a single spring constant, and choose units such that this constant (per unit volume) is unity. The resultant energy comes from the sum of the squared spring extensions, integrated across the bulk of the system:

\begin{align}
\label{eq:energy}
\en &= \frac{1}{2}\int d^\dim \rvec \, e_m(\rvec)e_m(\rvec).
\end{align}

\noindent Given the dependence of the spring extensions on both strains and gradients [(Eq.~(\ref{eq:ext})] each spring results in four energetic terms.
 Following a number of manipulations (see Supplementary Material) the two terms that break spatial inversion symmetry may be expressed as a total divergence term and hence reduced purely to a \emph{surface} contribution $\en_s$ to the energy expressed in terms of the outward-facing surface normal $\nvec$:

\begin{align}
\label{eq:surfacenergy}
\en_s = \frac{1}{2}\sum_m \int_{\textrm{surface}} d^{\dim-1} \rvec \left( \pvec_m \cdot \nvec\right) \frac{\bvecc^m_i \bvecc^m_j \bvecc^m_k \bvecc^m_l}
{|\bvec^m|^2} \strain_{ij} \strain_{kl}.
\end{align}

\noindent These terms therefore do not affect the bulk physics, but do modify the response of boundaries and interfaces, analogous to total gradient terms which are not present in nematics but do occur in cholesteric liquid crystals~\cite{meiboom1981theory}.
In addition to this surface term, we have spatial inversion symmetric terms due to the conventional strains and their gradients:

\begin{multline}
\label{eq:bulkenergy}
\en_b = \frac{1}{2} \sum_m \int d^d \rvec \,  
\frac{b^m_i b^m_j b^m_k b^m_l}{|\bvec^m|^2}  \times
\\
\Big( \epij \epkl + (\pvec_m \cdot \nabla) \epij \: (\pvec_m \cdot \nabla) \epkl \Big) 
\end{multline}

\noindent Hence, we recognize that the energy consists of three types of terms: conventional bulk strain energies leading to elasticities of the sort described in~\cite{lubensky2015phonons}, surface strain terms, and higher-order bulk terms consisting of strain gradients. Expanding the initial rigidity map further would generate higher-order bulk and surface terms. The bulk terms are even under spatial inversion and the boundary terms odd.

\begin{figure}[h]
\center
\includegraphics[scale=0.7]{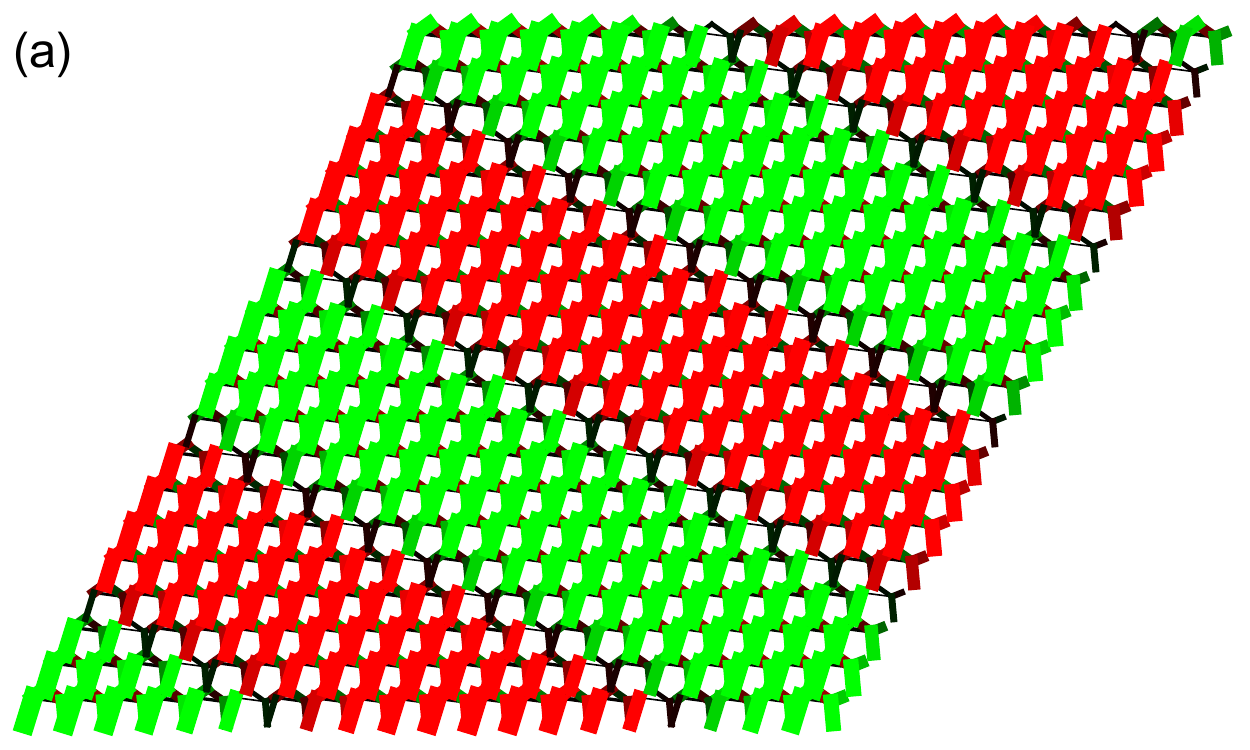}
\includegraphics[scale=0.7]{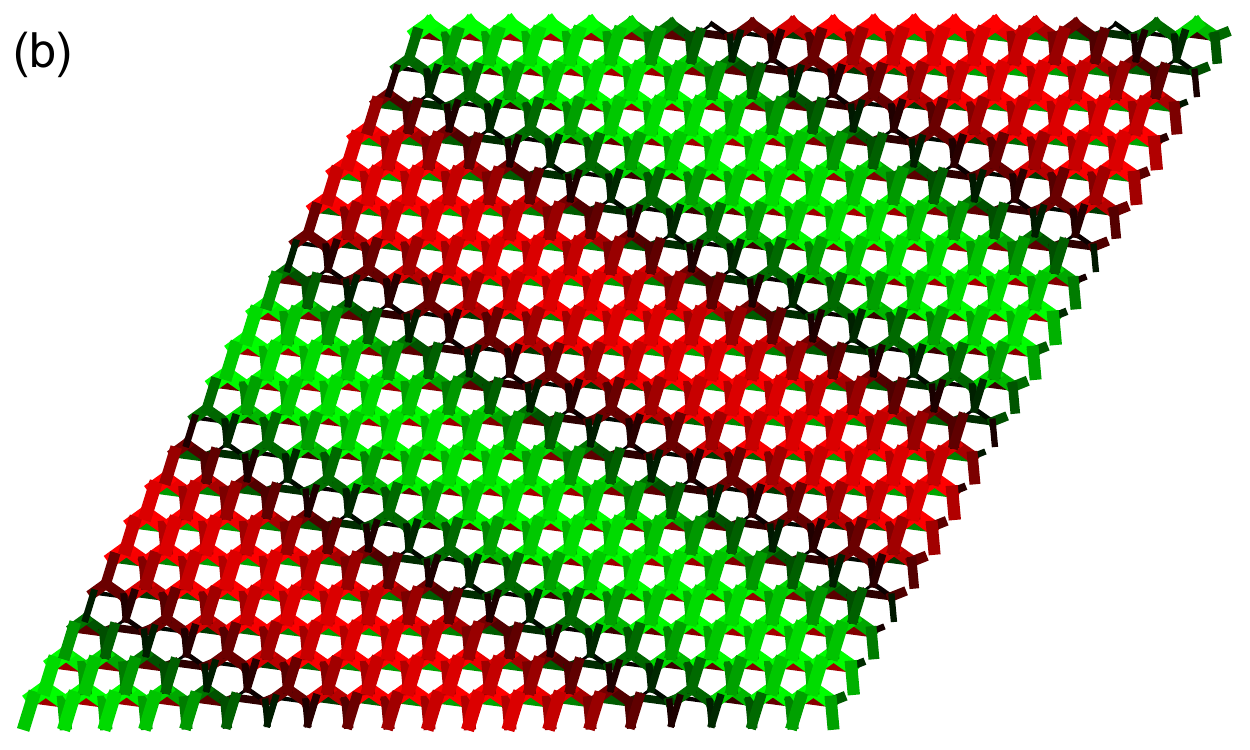}
\caption{(a) On a periodic system, we apply a particular strain $\epsilon_{ij} (\rvec) = (\epsilon_{xx}, \epsilon_{yy}) e^{i (\qvec \cdot \rvec)}$. This causes bonds to stretch (green) or compress (red).
(b) The system then relaxes by projecting onto space of self-stresses that mostly capture particles displacing over short distances but reducing the energy cost overall.}
\label{fig:relaxation}
\end{figure}

Note that we have not yet made any assumptions about either mechanical equilibrium or mechanical criticality. As such, even when the conventional elastic force balance is achieved (i.e., the divergence of the stress vanishes in the bulk) there are unbalanced forces on the individual degrees of freedom shown in Fig.~\ref{fig:relaxation}(a). Before we rectify this, let us illustrate the bulk and boundary energies with a minimal example, occurring in one dimension and with a single strain component $\epsilon(x)$. Let the extension of a bond be given by $e = (a-b\partial)\epsilon$, resulting in an energy

\begin{align}
\label{eq:exenergy}
E = \frac{1}{2}\int d x \, \left(a \epsilon - b \epsilon'\right)^2.
\end{align}

\noindent Requiring force balance in the bulk (namely, that the differential of the energy with respect to the strain function vanish) generates strain profiles exponentially localized to the two edges:

\begin{align}
\epsilon(x) = C_L \exp\left(-|a/b|x\right)+C_R \exp\left(+|a/b|x\right).
\end{align}

\noindent which are independent of the surface term $-ab \epsilon^2$, as indeed force-balanced terms must be. However, the surface term does determine the energy of these modes, such that depending on the sign of $a/b$ either the left-hand or the right-hand term costs zero energy. This is the linear limit of the nonlinear soliton chain considered in~\cite{chen2014nonlinear}. In this way, we see that the surface term determines the existence and position of zero-energy boundary modes. The above equations also reveal that, in general, this surface term and hence the zero-energy modes are independent of the bulk \emph{energetics}. Therefore, in the following sections we return to a general form of the bulk \emph{constraints} in order to consider the effects of mechanical equilibrium, mechanical criticality and topological protection. Note finally that the energy of Eq.~(\ref{eq:exenergy}) is characteristic of a mechanically critical lattice; more generally a system could have additional energetic terms such that neither surface mode cost zero energy.

\section{Mechanical relaxation and equilibrium in the microstructure}
\label{sec:relaxation}

From the energy functional of Eq.~(\ref{eq:energy}), we may obtain the  stress tensor $\stress_{ij}(\rvec) = \delta \en/\delta \strain_{ij}(\rvec)$ with the details of the functional differentiation described in the Supplementary Material. Just as each bond extension was obtained as a linear operator on the strain, now the stress is obtained as a linear operator on the bond extensions, which we denote as the initial equilibrium map $\eqc^0_{ij,\sind}$, again in analogy to the equilibrium matrix of lattice theories:

\begin{align}
\label{eq:stress}
\stress_{ij}(\rvec) &= \eqc_{ij,\sind}^0 \ext_\sind(\rvec), \\
\eqc_{ij,\sind}^0 & \equiv \frac{\bvecc_i^m \bvecc_j^m}{|\bvec|} \left[ 1-\sposc_k \partial_k \right].
\end{align}

The above equations describe the elastic relationships between stress, strain, bond extensions and energy. Note that these results mirror those of the lattice theories~\cite{lubensky2015phonons}. In particular, if we consider modes with spatial variation $\exp(i \qvec \cdot \rvec)$, the equilibrium map is simply the transpose of the rigidity map at the opposing wavevector:

\begin{align}
\label{eq:transpose}
\eqc_{ij,\sind}^0(\qvec) = \rigc_{\sind,ij}^0(-\qvec).
\end{align}

\noindent However, the smooth strain fields assumed here are unrealistic for isostatic systems, which undergo short-wavelength non-affine relaxation events to dramatically lower their energy in response to unbalanced forces within the unit cell.

Consider, as we have above, the extensions of bonds that occur when a strain field is imposed externally. These extensions generate tensions, which in turn generate unbalanced forces on the sites. These result in additional displacements of the sites, achieving force balance and energy minimization. These displacements may of course be obtained by solving whatever microscopic force-balance equations are appropriate to a given microstructure, but a more elegant and efficient method exists for systems close to the isostatic point. For such systems, there are only a few \emph{states of self stress}~\cite{calladine1978buckminster,kane2014topological}, which are sets of tensions that generate no force on any site. As has been shown~\cite{mao2018maxwell}, the post-relaxation bond extensions $\evec$ are precisely the projection of the pre-relaxation bond extensions $\evec_0$ into the space spanned by an orthonormal set of states of self stress, $\{\svec_i\}$. 

For Maxwell lattices, which have equal numbers of site degrees of freedom and bond constraints, this technique has proven useful to obtain the uniform elastic response. Such a lattice has $d$ modes of translation that result, via the Maxwell-Calladine index theorem, in $\dim$ states of self stress. The $\dim(\dim+1)/2$ components of the elasticity tensor are all obtained from these, implying that $\dim(\dim-1)/2$ strains cost zero energy. These are known as Guest modes \cite{guest2003determinacy}. This method, which accounts naturally for the large intra-cell relaxation, yields the correct moduli, whereas assuming a uniform strain field overestimates them substantially.

In the present analysis, we wish to consider external fields that are applied over length scales large compared to the unit cell, such that displacement fields necessarily vary smoothly over these scales. At the same time, the fields may vary dramatically \emph{within} the cell, as when, e.g., a rigid triangular unit undergoes a rotation. To that end, we wish to consider the behavior of structures such that the center of mass of each cell is coupled to external fields (such as induced by smooth, rigid mechanical barriers) while the other, internal degrees of freedom (rotations and shears) are allowed to relax and achieve mechanical equilibrium. In this way, we retain a theory expressing the energetics in terms of the slowly-varying strain field across the periodic structure, while still permitting deformations that relax energy within the crystal cell.

To that end, we identify the smooth displacement (alternately strain) fields $\uvec(\rvec)= \uvec \exp \left( i \qvec \cdot \rvec \right)$. These are the fields we wish to retain, while the short-range relaxations within the unit cell are allowed to occur to lower the system energy. Note that although $\uvec$ has only $\dim$ components, it captures all of the allowed strains. For $\qvec \ne 0$, there exist  $\dim (\dim-1)/2$ mechanical compatibility conditions~\cite{guest2003determinacy} that are satisfied by the $\dim (\dim+1)/2$ components of the strain, such that they do indeed derive from a valid displacement field defined by $\dim$ components. 

How, then, do we allow relaxation of the remaining degrees of freedom? Previously, we had required that all the components of the force $\fvec = \eq \evec$ on the sites resulting from the extensions must vanish. Now, though, we wish to allow forces on the smooth strain fields to remain unbalanced. 
For example, if we consider a set of strain fields that generates a force on the center of mass of a cell and a torque on one of its rigid elements, we would allow the cell to relax to alleviate the torque while retaining the center of mass force.
That is, if $\{|v_i\rangle\}$ are the set of externally imposed center-of-mass displacements on the cells
 we wish to allow no center-of-mass forces, such that we allow $\langle v_i | f \rangle \ne 0$ but require that the other components of the forces vanish. To that end, our self stresses are now defined as an orthonormal basis for the nullspace of the equilibrium matrix with these modes projected out:

\begin{align} 
\label{eq:newnull}
\{ |s_j \rangle \} = \textrm{Null} (\mathbb{1} - |v_i \rangle \langle v_i | )\eq.
\end{align}

For the periodic systems currently under consideration, modes at different wavevectors are orthogonal. Consequently, we can consider rigidity and equilibrium maps at a particular wavevector, such that $\partial_j \rightarrow i \qvecc_j$ and a strain at a given wavevector results in spring extensions only at the same wavevector. Consequently, there are only $\dim$ modes at any given wavevector that need be projected out in Eq.~(\ref{eq:newnull}). Each such mode consists of displacements of sites along one of the Cartesian directions, with spatial dependence $\exp(i \qvec \cdot \rvec)$. The result is $\dim$ states of self stress.

We thus obtain the primary object under consideration in the present work, the \emph{rigidity map} $\rigc_{m,ij}$, which describes the extensions which result from a smooth strain field following the short-distance relaxations within the unit cell. From this, the equilibrium map may be derived in precisely the same way as the initial equilibrium map was derived from the initial rigidity map. In terms of the initial rigidity map of Eq.~(\ref{eq:rig}) and $s_{m,n}$, the $n^\textrm{th}$ component of the $m^\textrm{th}$ state of self stress,

\begin{align}
\label{eq:rigtensor}
\rigc_{\sind,ij}(\qvec) =  s^*_{m,n} \rig^0_{n,ij}(\qvec), \\
\eqc_{ij,\sind}(\qvec) =  \eq^0_{ij,n}(\qvec) s_{m,n}.
\end{align}

The above analysis implies that, if we think of the rigidity map as a matrix acting on the $\dim$ independent components of the strain, it similarly results in $\dim$ independent self stresses, and is thus a square matrix. That is precisely the Maxwell condition for lattice theories, that the energetic constraints and the degrees of freedom are equal in number. Thus, we arrive at a natural condition to extend the Maxwell condition to theories of continuum fields: that the configuration spaces and constraints be equal in number.

A final property proves crucial to the low-energy response of the lattice. The aforementioned Guest modes ensure that the rigidity maps have zero modes at low wavevectors. That is, generically the eigenvalues of the rigidity map would have $O(\qvecc^0)$ components, yet for the Guest mode the leading contribution would be $O(\qvecc^1)$.

\section{Topological polarization in the continuum}
\label{sec:topology}

We have now generated a continuum map that describes the relationship between local strain degrees of freedom and a like number of energetic constraints, retaining the Maxwell criterion of the lattice theories. However, it is not a priori clear that this formulation captures the phenomena of the lattice theory, including topological protection of modes at interfaces, edges and defects. Such modes are generated and protected by topological invariants defined by the homotopy class of loops across the Brillouin Zone, which no longer exists in our continuum formulation. Do the zero modes have continuum descriptions, and do they retain topological protection? We resolve these questions in the affirmative.

Since we mean to include edge modes, we allow $\qvec$ to be complex, with the imaginary part representing the rate at which the mode decays away from the edge. In particular, suppose we have an edge in the second spatial direction, on the line $(0,\rvecc_y)$, with the system extending to its right, with $\rvecc_x>0$. A mode extending along and exponentially localized to this edge would have $\textrm{Im}(\qvecc_y)=0$ real and $\textrm{Im}(\qvecc_x)>0$, while one on the opposite edge would have $\textrm{Im}(\qvecc_x)<0$.

Because any extensions of bonds cost energy, zero-energy modes are precisely those that lie in the 
 nullspace of the rigidity map of Eq.~(\ref{eq:rigtensor}). Note that the use of this map means that we are allowing the local structure to relax to minimize energy, in this case to zero. Thinking of this map as a square matrix, we may take its determinant, $\det(\qvec)$, which vanishes if and only if a zero mode exists. Because of the two modes of translation, $\det(\qvec)$ has  a double zero at $\qvec = 0$ and hence takes the following form:

\begin{align}
\label{eq:detxy}
A_{2,0} \qvecc_x^2 + A_{1,1} \qvecc_x \qvecc_y + A_{0,2} \qvecc_y^2 + i A_{3,0} \qvecc_x^3 + i A_{2,1}\qvecc_x^2 \qvecc_y + \ldots,
\end{align}

\noindent with $A_{n_1 n_2}$ real coefficients set by the microstructure. 
Because our rigidity map has units, $A_{ij} q_x^{n_1} q_y^{n_2}$ has units of the volume of $d$-dimensional space (two, here).
Terminating the expansion to order $n$ in $\qvecc$ indicates the presence of $n$ zero modes. Of these, two take the form

\begin{align}
\label{eq:alpha}
\qvecc_x = \alpha_\pm \qvecc_y + i \bcoef_\pm \qvecc_y^2,
\end{align}

\noindent while the others are short-wavelength modes of order $\qvecc_y^0$. $\alpha$ is determined by the coefficients of order $O(\qvecc^2)$, 
while the signed inverse decay length $\invd_\pm \equiv \bcoef_\pm \qvecc_y^2$ depends on the second and third order terms of Eq.~(\ref{eq:detxy}). While a lattice may have short-wavelength modes, our continuum formulation deliberately excludes them and hence the short-wavelength modes here are non-physical and dictated by the order at which we terminate our expansion. Our long-wavelength zero edge modes are then restricted to wavevectors obeying the above equation. The trivial state is the one in which  $\invd_\pm$ take opposing signs, whereas in a polarized state both $\invd_\pm$ would take the same sign. Thus, to capture the edge mode polarization, a topological invariant must count the numbers of positive and/or negative $\invd_\pm$.

Because the loss of periodicity destroys the Brillouin Zone, a natural approach would be to perform an integral instead over all real values of $\qvecc_y$. This could lead, via the extended real number line, to a non-contractible loop such as one has in the periodic case, analogous to an approach that was applied in \cite{shankar2017topological,souslov2017topological} to determine Chern numbers on the surfaces of spheres rather than tori. However, this procedure would not be appropriate here since it would include the fictitious short-wavelength modes. 

Instead, to capture the long-wavelength behavior, we introduce $|\lvec_1|$, the length of the first lattice primitive vector, and consider edge modes for which the transverse component of the wavevector is small: $\qvecc_y |\lvec_1| \ll 1$. In selecting modes over which to integrate, we wish to ensure that we consider values of the remaining component of the wavevector that extend far beyond those while also ignoring fictitious short-wavelength modes. That is, we wish to consider values of $\qvecc_x$ which satisfy $\qvecc_y |\lvec_1| \ll \qvecc_x |\lvec_1| \ll 1$. We choose then for $\epsilon \equiv \qvecc_y |\lvec_1|$ to integrate over values of $\qvecc_x |\lvec_1|$ extending to $\epsilon^{1/2}$. As described in detail in the Supplementary Material, there are alternate methods which may capture the transition more sharply. The resulting topological invariant describing the numbers of modes on the left and right edges is:

\begin{align}
\label{eq:invariant}
N_L - N_R = \frac{1}{\pi} \lim_{\epsilon \rightarrow 0^+} \int_{-\sqrt{\epsilon}/|\lvec_1|}^{\sqrt{\epsilon}/|\lvec_1|} d \qvecc_x \partial_{\qvecc_x} \textrm{arg} \,\det(\qvecc_x, \qvecc_y = \epsilon).
\end{align}

\begin{figure}[h]
\center
\includegraphics[scale=0.5]{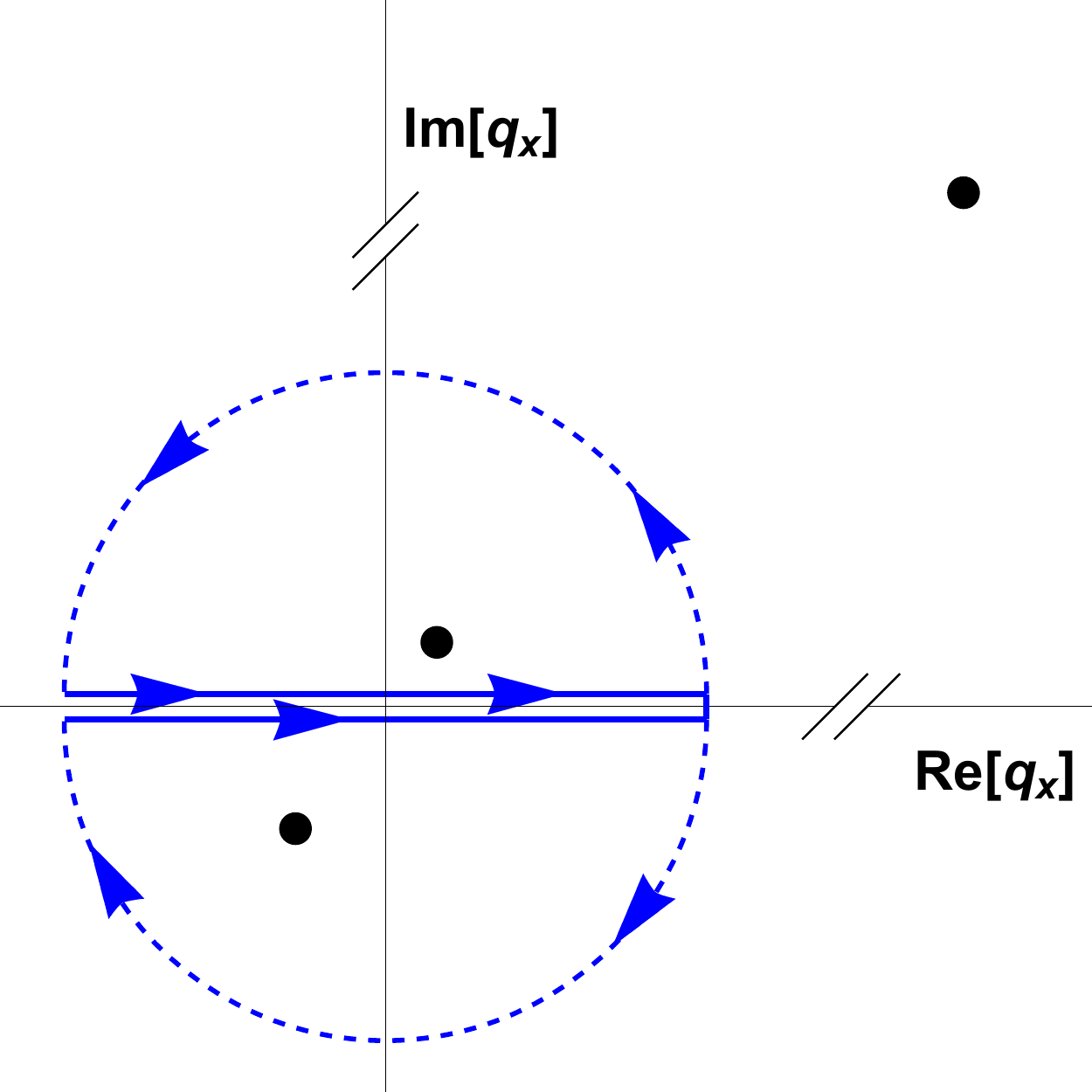}
\caption{The contour used to establish the relationship between the bulk structure and the locations of the zero modes. Via complex analysis, the full contour counts the difference in numbers of long-wavelength zero modes on the left and right edges while remaining insensitive to the short-wavelength modes. As described in the text, for this particular contour the dashed portions may be neglected, such that the locations of the zero modes are determined solely by the bulk physics captured in the solid lines.}
\label{fig:winding}
\end{figure}

\noindent This relationship follows from the contour in the complex plane shown in Fig.~\ref{fig:winding}. In that figure, we search for a relationship between the behavior of the bulk modes, shown on the real axis, and the zero-energy surface modes, represented as black points. Applying the Argument Principle means that the long-wavelength modes will cause the phase of the determinant of the rigidity map to wind by $\pm 2\pi$ for modes on the left (right) edges. The curved portions of the contour themselves have imaginary components of the wavevector and hence are boundary modes that cannot be considered in establishing a bulk-boundary correspondence, a key signature of topological protection. However, while the contributions of the curved portions are never small, their \emph{difference} does vanish in the given limit. As such, we may determine the presence of zero modes solely through the portions of the contour corresponding to bulk modes (solid lines). In this way, the topological invariant becomes quantized (integer) precisely in the long-wavelength limit, as shown in detail in the Supplementary Material. Indeed, this notion of the topological invariant emerging in the long-wavelength limit is implicit in other continuum treatments, insofar as they likewise consider length scales over which atomic details may be neglected. This continuum object in fact bears a closer relationship to conventional polarization than does the discrete analog, which counts the number of zero modes on a particular edge rather than the difference between the two edges.

\begin{figure}[h!]
\center
\includegraphics[scale=0.55]{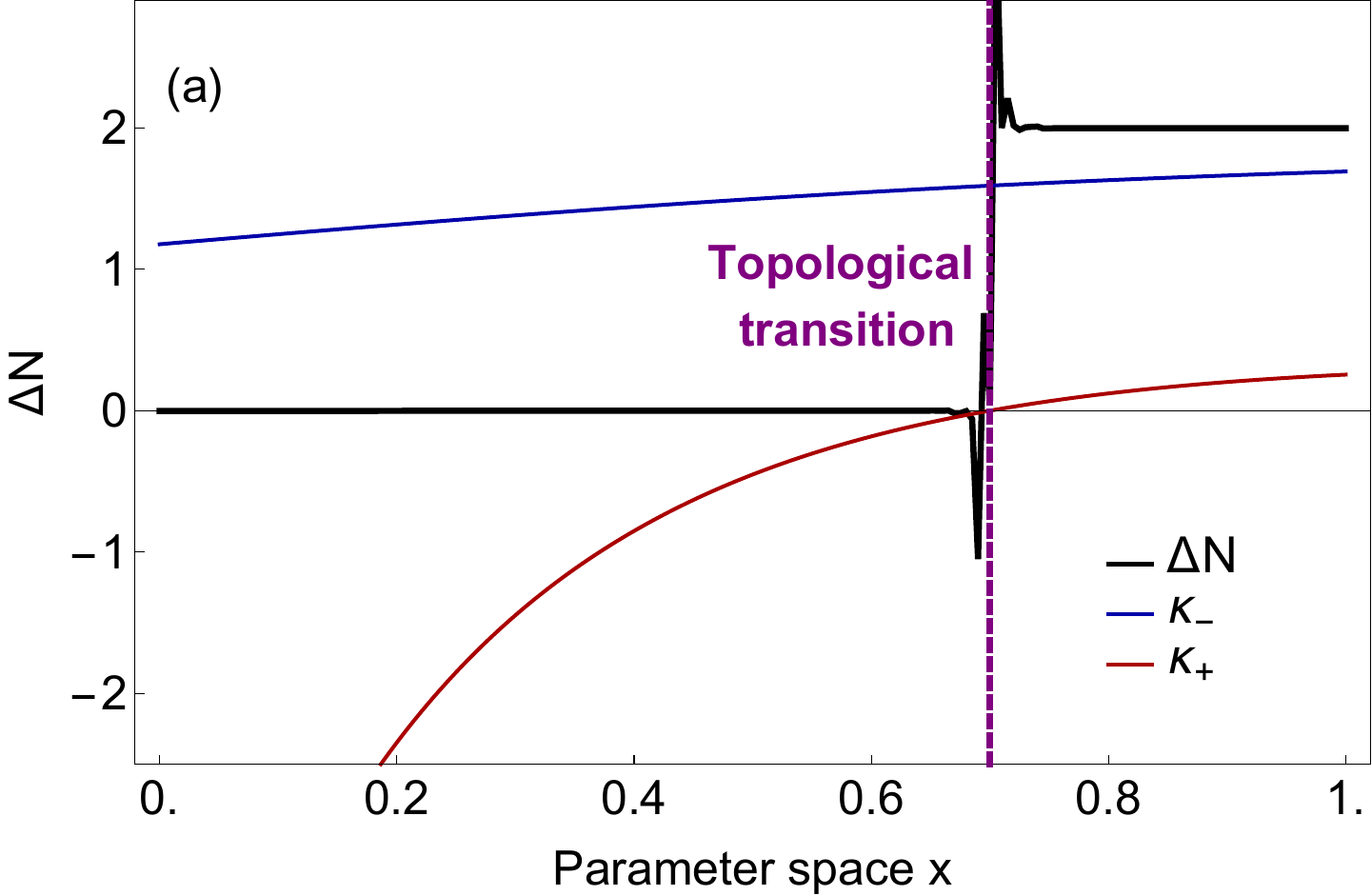}
\includegraphics[scale=0.40]{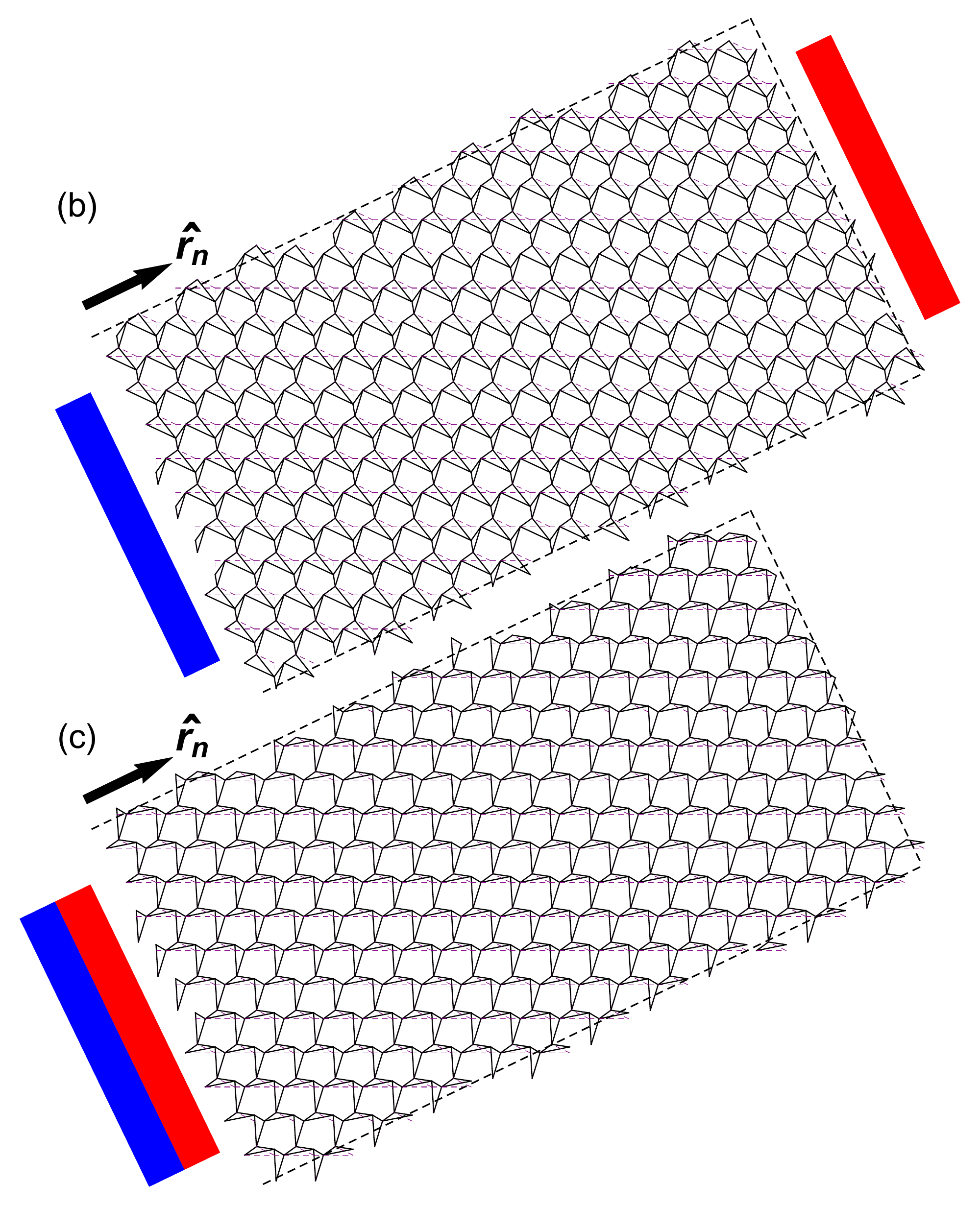}
\caption{(a) Topological transition as we deform the Kagome lattice. The geometry of the system is parametrized as $\gvec (x) = x \gvec_1 + (1-x) \gvec_2$ where $\gvec_1, \gvec_2$ are geometric configurations of two Kagome systems with respective topological polarizations $0$ and $2$ along the $\rh_n$ direction. $\kappa_\pm$ are the signed inverse decay lengths, such that when one vanishes the associated mode lies in the bulk and when both have the same sign the system is polarized. (b) The system in the $\gvec_1$-configuration with trivial polarization: there is a zero mode on the left side of the considered orientation $\rh_n$ and one on its opposite side. (c) The system in non-trivial polarization ($\Delta N = 2$): this time, the two modes are on the left side of the direction. On all three figures, the purple dashed lines represents the shape of the system at the topological transition (when $x=0.7$).} 
\label{fig:poltransition}
\end{figure}

To demonstrate that nonzero topological polarizations do occur, we consider a class of generalized kagome lattices, consisting of a periodic cell with three sites in two dimensions, joined by three intracellular and three intercellular bonds as shown in Fig.~\ref{fig:poltransition}. In that figure, we consider a 1D family of such lattices that undergoes a topological transition at which deformation of the lattice shifts a boundary mode from one edge to the opposing edge, polarizing the system. As shown in Fig.~\ref{fig:poltransition}(a), our numerical technique generates noticeable error as described in the Supplementary Material very close to the transition point, which can also be identified by direct geometrical means.

\subsection{Topological polarization as a vector}

\begin{figure}[h]
\center
\includegraphics[scale=0.55]{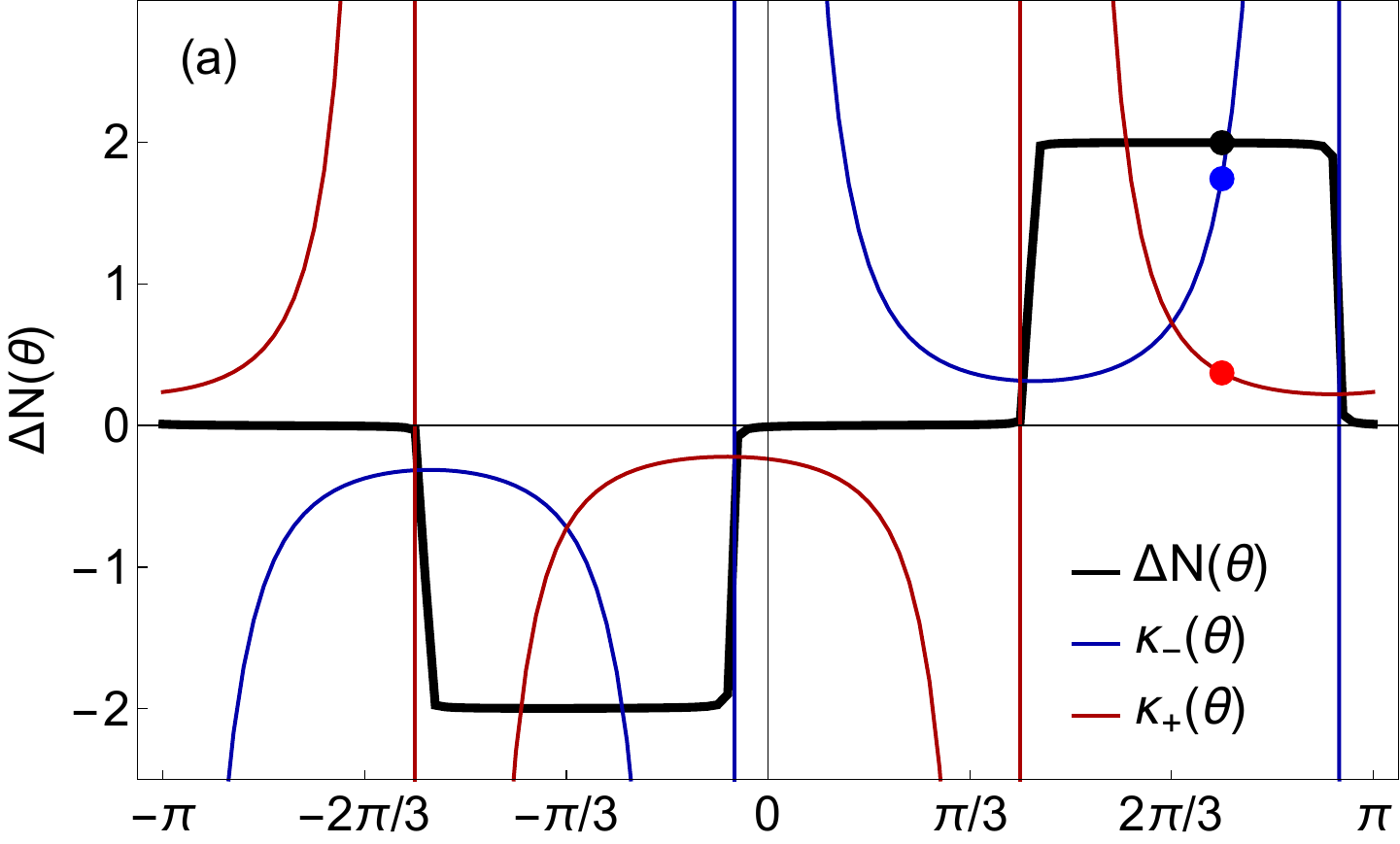}
\includegraphics[scale=0.40]{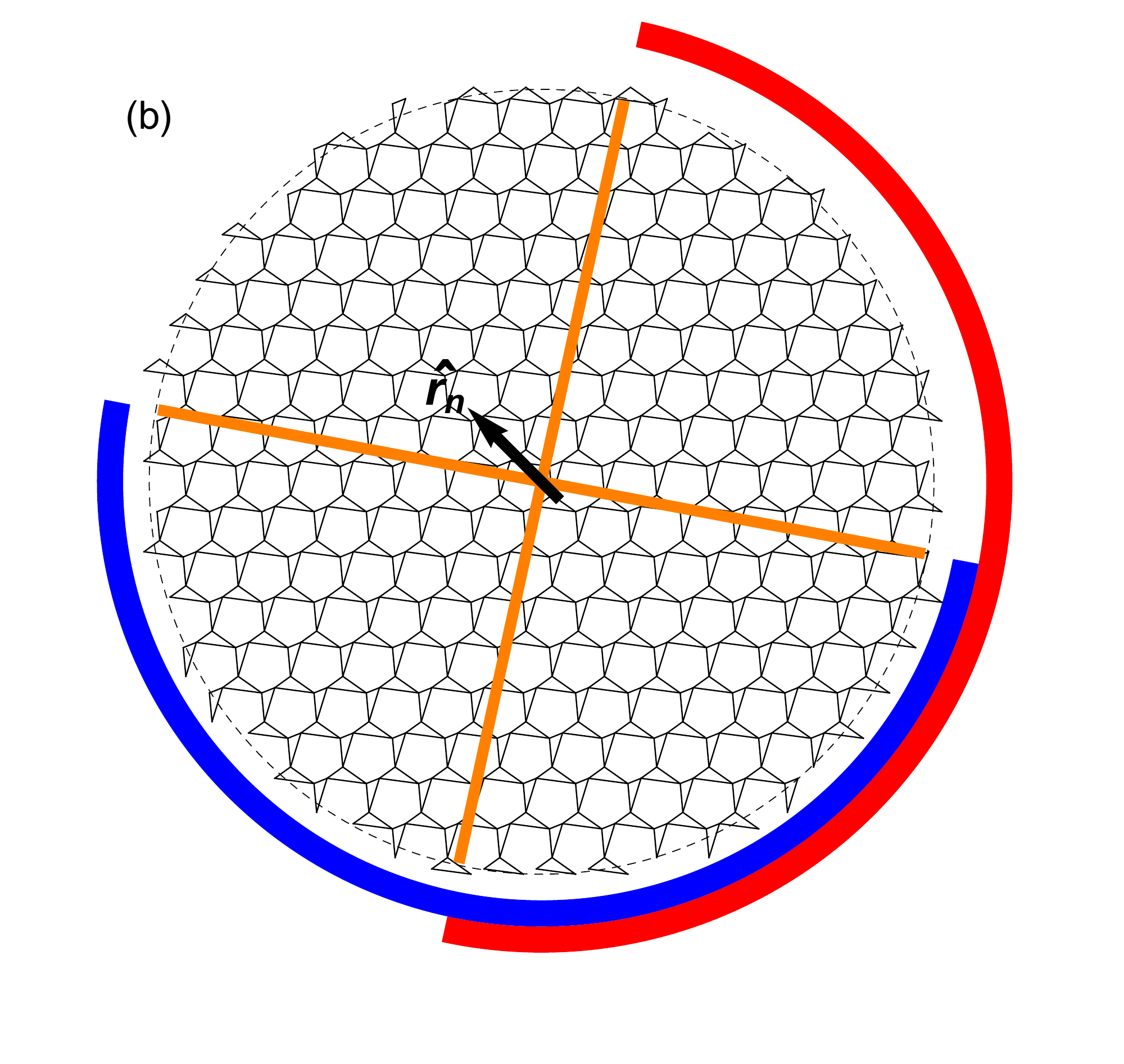}
\caption{(a) For a fixed system, we numerically compute the signed inverse decay lengths of each mode ($\invd_- (\ang), \invd_+ (\ang)$) and the topological polarization $\Delta N(\ang)$ as a function of the normal direction $\rh_n (\ang)$. (b) The polarization changes as the considered direction crosses either soft directions (orange lines). One can then find the regions of the lattice where the $+/-$ edge modes are located (blue/red arcs).} 
\label{fig:polrotation}
\end{figure}

In the continuum, it becomes natural to ask about the polarization not only at a certain interface --- for now, we've focused on a vertical interface, with decay along the horizontal direction ($x$-direction) --- but for every possible interface. Specifically, we look for $\Delta N (\ang)$, the difference between the number of zero modes on the ``left'' and ``right'' edges when the vector pointing from the left to the right edge is $\hat{q}_n(\ang) \equiv \cos \ang \, \hat{x} + \sin \ang \, \hat{y}$. In this way, we can recast the wavevector in terms of the components along this normal direction, $\qvecc_n \equiv \qh_n \cdot \qvec$ and the component $\qvecc_t$ along the tangent direction $\hat{q}_t(\ang) \equiv -\sin \ang \, \hat{x} + \cos \ang \, \hat{y}$. This allows us to re-write the determinant in the new $(\qvecc_n, \qvecc_t)$ basis and find the long-wavelength modes of present interest:
\begin{align}
\det (\qvecc_n, \qvecc_t) = \det \big(\qvecc_x = \cos \ang \qvecc_n - \sin \ang \qvecc_t, \nonumber \\
\qquad \qquad \qvecc_y = \sin \ang \qvecc_n + \cos \ang \qvecc_t \big)
\nonumber \\
= A'_{2,0} \qvecc_n^2 + A'_{1,1} \qvecc_n \qvecc_t + A'_{0,2} \qvecc_t^2 + i A'_{3,0} \qvecc_n^3 + \ldots
\\
\implies q_n = \alpha_\pm (\ang) \qvecc_t + i \invd_\pm (\ang) \qvecc_t^2,
\label{eq:detnt}
\end{align}

\noindent where once again the sign of  $\invd_\pm (\ang) \equiv \bcoef(\ang) \qvecc_t^2$  determines the topology in the particular $\rh_n (\ang)$-direction. This new expression allows us to describe how the topological polarization changes at different angles. What we observe is that the modes flip from one side to the other as the direction $\rh_n (\ang)$ crosses a particular set of directions (indicated by orange lines on Fig.~\ref{fig:polrotation}). Those regions are called soft directions and we address their nature in the following section.

\subsection{Soft Directions}

Focusing on the long-wavelength zero modes established in Eq.~(\ref{eq:alpha}), we find the existence of soft directions (indicated by orange lines on Fig.~\ref{fig:polrotation}(b)), characteristic of the lattice~\cite{rocklin2016mechanical} and determined by the value of the $\alpha_\pm$ coefficients:
\begin{align}
\qvecc_x = \alpha_\pm \qvecc_y \implies \hat{\qvecc}_\pm = \frac{(\alpha_\pm, 1)}{\sqrt{1+\alpha_\pm}}.
\label{eq:soft}
\end{align}

\noindent We can then re-write the determinant of a generic wave-vector $\qvec$ in this new basis $\qvec = \qvecc_+ \hat{\qvecc}_+ + \qvecc_- \hat{\qvecc}_-$:
\begin{align}
\det (\qvec) = A''_{1,1} \qvecc_+ \qvecc_- + i A''_{3,0} \qvecc_+^3 + i A''_{2,1}\qvecc_+^2 \qvecc_- + \ldots,
\label{eq:detsoft}
\end{align}

\noindent so that if a wavevector lies exactly on either soft direction (i.e $\qvec = \qvecc_+ \hat{\qvecc}_+$ or $= \qvecc_- \hat{\qvecc}_-$), it is by definition a zero of our \emph{rigidity map} at least to 2nd order.

\begin{figure}[h]
\center
\includegraphics[scale=0.3]{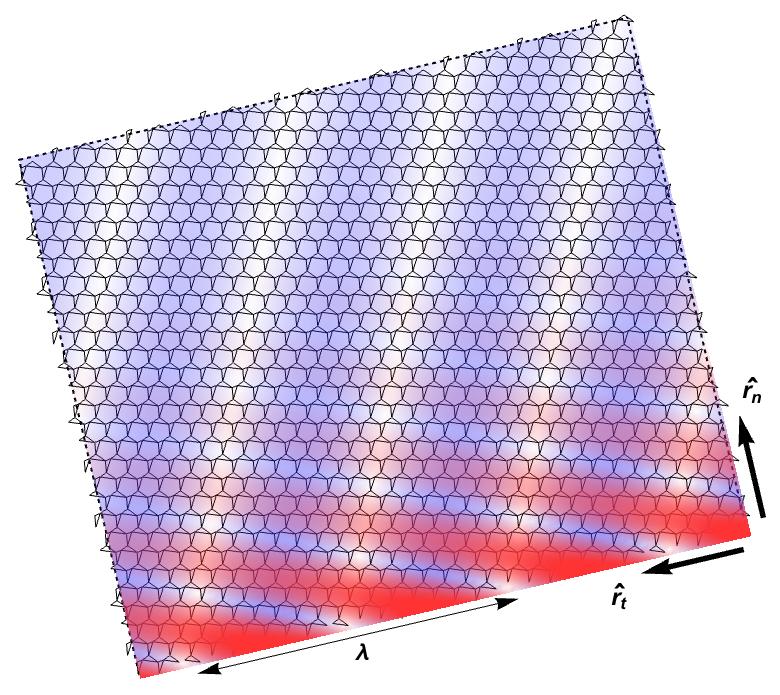}
\caption{Shape of the edge modes in a system with polarization $\Delta N = 2$. We impose a mode with wavelength $\lambda = 15.6 \, |\lvec_1|$ on the boundary $\rh_t$ (where $\lvec_1$ is the first lattice primitive vector). This mode then propagates through the bulk and decays along the direction $\rh_n$. Each mode (red, blue) varies sinusoidally along its corresponding soft direction and therefore also varies along the boundary with a longer wavelength. The modes decay into the bulk over length scales much longer than this wavelength.}
\label{fig:edgemode}
\end{figure}

\section{Experimental length and energy scales}
\label{sec:length}

The triumph of conventional elasticity is its ability to capture mechanical response at length scales extending to system size, far beyond those of the underlying, unobserved atomic interactions. It is not clear a priori whether the boundary modes which we consider extend to macroscale systems or whether they are valid only on the ``atomic'' length scales of the unit cell, those already captured by the lattice theory. Indeed, the dramatic change in edge stiffness predicted by simple central-force models~\cite{rocklin2017transformable} have resulted in relatively modest differential stiffnesses in 3D-printed systems~\cite{bilal2017intrinsically}.
In fact, as we show here, the boundary modes extend to wavelengths intermediate between the unit cell and the system size, establishing new criteria for experimentally realizing strong topological effects.

Let us consider in particular the imposition of a distortion on a boundary with wavenumber $\qvecc_t$ on an interface with tangent direction $\rh_t (\ang)$. In order to minimize energy, the system will undergo a distortion associated with one (or both) of the two soft directions described in the preceding section. We thus search the space of zero modes described in Eq.~(\ref{eq:detsoft}) for a zero mode with the appropriate component along the surface tangent direction, resulting in a wavevector of the form

\begin{align}
\qvec_\pm = \frac{\qvecc_t }{\hat{\qvecc}_{\pm} \cdot \hat{\qvecc}_t} \hat{\qvecc}_+\: 
\pm \: i \qvecc_t^2 \frac{A''_{3,0(0,3)}}{A''_{1,1}} \, 
\frac{| \hat{\qvecc}_+ , \hat{\qvecc}_- |}{(\hat{\qvecc}_\pm \cdot \hat{\qvecc}_t)^3} \hat{\qvecc}_n, 
\label{eq:zmsoft}
\end{align}
 
\noindent where $| \hat{\qvecc}_+ , \hat{\qvecc}_- |$ denotes the determinant of the matrix of the given columns. Expressed in terms of the angles $\theta_\pm, \ang$ of the soft directions and the normal direction, this is

\begin{align}
\qvec_\pm &= \frac{\qvecc_t }{\sin(\theta_\pm - \ang)} \, \qh_\pm \,
\pm \, i \qvecc_t^2 \,  \frac{A''_{3,0(0,3)}}{A''_{1,1}} \, \frac{\sin (\theta_+ - \theta_-)}{\sin^3 (\ang - \theta_\pm)} \qh_n.
\label{eq:fullzeromode}
\end{align}

\noindent  We note that the decay part of this expression includes parameters of the systems that come from the determinant of our \emph{rigidity} mapping, i.e. the $A''$ terms of equation (\ref{eq:fullzeromode}), and hence can't be measured by the bulk response.

We now wish to express this relationship in terms of unitless ratios (in brackets) between the quantities with units of the lattice length scale, the decay length and the wavelength of the surface distortion, $\lambda = 2 \pi/q_t$:

\begin{align}
\left[\frac{\dl_\pm}{|\lvec_1|}\right] &= \pm \frac{1}{(2\pi)^2}\left[\frac{\lambda}{ |\lvec_1|} \right]^2 \left[\frac{|\lvec_1| A''_{1,1}}{A''_{3,0 (0,3)}}\right] \frac{\sin^3 (\ang - \theta_\pm)}{\sin (\theta_+ - \theta_-)}.
\label{eq:decaylength}
\end{align}

\noindent From the above expression, we see that the number of cells over which the topological mode decays is generically on the order of the square of the number of cells over which it extends on the surface. For example, if the wavelength extends over thousands of unit cells, the boundary mode generically decays on the scale of millions of cells. Hence, the surface theory extends from microscopic wavelengths up to the order of the geometric mean of the system size and unit cell size. However, this geometric relationship is strongly modified by the direction of the interface. When the soft direction is nearly aligned with the normal direction, the wavelength of the modes in the bulk becomes much smaller than that on the boundary, and the decay length is correspondingly reduced, meaning that the effect may be observed in smaller systems.

Beyond these concerns of length scale, the surface theory also relies on appropriate energy scales. While a full energetic analysis is beyond the scope of the present work, consideration of microscopic interactions and length scales permits us to estimate additional criteria to observe continuum topological polarization. Real systems contain additional constraints beyond those included in the Maxwell rigidity map. In a 3D-printed frame, in additional to strong energetic costs to central-force compressions and extensions, beams possess finite bending stiffness not found in central-force springs. In our topological modes, these bending costs are imposed on a length scale proportional to the square of the wavelength, whereas a conventional Rayleigh-type surface mode would impose greater central-force energy costs over a smaller volume, extending to a depth proportional only to wavelength. Hence, the topological surface modes should only be strictly observable on wavelengths sufficiently short that they lie below the conventional modes in energy.

\section{Conclusions}
\label{sec:conclusions}

We have presented a new method for describing the energetic effects of imposed macroscopic strain fields and their gradients on a microstructure that undergoes microscopic relaxation. We arrive at separate expressions for bulk and surface energies, and establish for critically-coordinated systems an elastic \emph{bulk-boundary correspondence} between the bulk structure and topologically protected zero-energy modes. The underlying topological invariant establishes the relative numbers of zero modes on two opposing boundaries, as either number separately would rely on short-wavelength physics. This establishes a sort of mesoscale elasticity, with topological surface modes existing on length scales far beyond the unit cell but necessarily far below the system size. Notably, this may lead to experimental demonstrations of topological polarization based on macroscale strains even when microscopic structure remains unobserved.
The reader is encouraged to compare our approach with the illuminating~\cite{sun2019universal}, which was posted independently during the preparation of this manuscript.

These extensions of the lattice theory to the continuum present several avenues for further study. Lattice theories have considered bulk response~\cite{rocklin2017directional}, topological defects~\cite{paulose2015topological}, buckling failure~\cite{paulose2015selective} and fracture~\cite{zhang2018fracturing}. All of these phenomena may extend to the continuum in intriguing ways. Following the realization of topological states in kirigami sheets~\cite{chen2016topological}, incorporating curvature into our continuum model may yield new physics. 

Intriguingly, recent work has also shown that topological boundary modes can arise even in amorphous systems due to mesoscale structure~\cite{mitchell2018amorphous}. Similarly, our surface theory relies on the short-length  structure, suggesting that it may extend even to non-periodic systems, though topological polarization still relies on breaking spatial inversion symmetry. If indeed such local structure can describe directional response, it may underlie artificial structures which program intricate mechanical responses~\cite{rocks2017designing,yan2017architecture,KimBas2019}, themselves inspired by biological allostery.

Nonlinearities too may prove tractable in the general continuum theory, as when they lead to topologically protected solitons in one-dimensional systems~\cite{chen2014nonlinear}. 

In the present work, we considered a critical balance between constraints and independent components of the strain field that was derived by the critical coordination of the microscopic system. It is an open question as to whether such theories may emerge in the continuum without being present in the microscopic system. Structures such as those consisting of rigid square pieces joined at corners have a nonlinear zero-energy dilation mode~\cite{grima2000auxetic} and are well-described by continuum theories~\cite{coulais2018characteristic, deng2019focusing}. However, such systems rely on symmetry to achieve their deformation mode and do not have zero-energy boundary modes.

\emph{Acknowledgments:} The authors gratefully acknowledge helpful conversations with Danilo Liarte and Jim Sethna. Part of this work was supported by the Bethe/KIC Fellowship and the National Science Foundation through Grant No. NSF DMR-1308089 (D. Z. R.).

\bibliography{ctbib}

\newpage

\begin{appendices}

\section{Deriving the equilibrium map $\Qijm$ from the rigidity map $\Rmij$}
\label{secsi:qr}

We expect the components of the stress tensor to be linear in the spring tensions $e_m$, and we define the \emph{equilibrium map} as
\begin{align}
\label{eqsi:eq}
\sij (\rvec) = - \Qijm \, e_m (\rvec).
\end{align}

\noindent The following analysis can be made on either pre/post relaxation mappings. Given that the bonds' extensions are linear in the components of the strains
\begin{align}
\label{eqsi:rig}
e_m(\rvec) = \Rmij \epij (\rvec),
\end{align}

\noindent the total energy of the system is:
\begin{align}
E = \frac{1}{2} \int d\rvec e_m^2 (\rvec) =  \frac{1}{2} \int d\rvec \big(\Rmij \epsilon_{ij} (\rvec) \big)^2.
\end{align}

\noindent Given that
\begin{subequations}
\begin{align}
\sigma_{ij} (\rvec) 
&= -\frac{\delta E}{\delta \epsilon_{ij} (\rvec)} 
\\
&= - \frac{\delta}{\delta \epsilon_{ij} (\rvec)}  \frac{1}{2} \int d\rvec \big(\Rmij \epsilon_{ij} (\rvec) \big)^2
\\
&= - \Rmij \: \Rmij \epsilon_{ij} (\rvec)
\\
\implies &\sigma_{ij} (\rvec) = - \Rmij \, e_m (\rvec),
\end{align}
\end{subequations}

\noindent where in the final line we used the linear relationship of Eq.~(\ref{eqsi:rig}). From Eq.~(\ref{eqsi:eq}) we recognize here the equilibrium mapping $\Qijm$, meaning
\begin{align}
\Qijm = \Rmij.
\end{align}

%%%%%%%%%%%%%%%%%%%
%%%%%%%%%%%%%%%%%%%

\section{Relating equilibrium and rigidity maps in reciprocal space with periodic structure}

In this section, we incorporate periodic structure into the relationships described in the previous section (Sec.~(\ref{secsi:qr})). As we will see, the dependence on wavevector is reversed for the equilibrium and rigidity maps, the key result first obtained for lattice structures which permits topological polarization. To that end, we first consider more general, nonlocal classes of rigidity and equilibrium maps than we have previously:
\begin{align}
e_m (\rvec) &= \int d\rvec' \Rmij (\rvec, \rvec', \partial') \epij (\rvec'), \\
\sij (\rvec) &= - \int d \rvec' \Qijm (\rvec, \rvec', \partial') e_m (\rvec').
\end{align}

\noindent where the inclusion of ``$\partial$'' indicates that the maps can involve gradients. Repeating the process above, we again obtain stresses from our energy functional. Now though, we apply the well-known result that such functional differentiation flips the sign of the gradients. This may be seen by explicitly writing out gradients of the delta functionals that are used in functional differentiation and integrating by parts. The result, in this case, is that

\begin{subequations}
\begin{align}
\Qijm (\rvec&, \rvec', \partial') = 
- \frac{\delta \sij (\rvec)}{\delta e_m (\rvec')} 
\\
&= \frac{\delta^2 E}{\delta \epij(\rvec) \, \delta e_m (\rvec')}.
\\
&= \frac{\delta}{\delta \epij(\rvec)} \left( \frac{\delta E}{\delta e_m (\rvec')} \right)
\\
&= \frac{\delta}{\delta \epij(\rvec)} \, e_m (\rvec') 
\\
&= \frac{\delta}{\delta \epij(\rvec)} \, \int d \rvec'' \Rmij (\rvec', \rvec'', \partial'') \epij (\rvec'')
\\
&= \int d \rvec'' \Rmij (\rvec', \rvec'', \partial'') \delta (\rvec'' - \rvec),
\end{align}
\end{subequations}

\noindent leading to the desired result in real space:
\begin{align}
\Qijm (\rvec, \rvec', \partial')  = \Rmij (\rvec', \rvec, -\partial) 
\end{align}

\noindent We now return to incorporating our system's translational invariance and (semi-)local interactions, such that $\Rmij (\rvec, \rvec', \partial) \rightarrow \Rmij (\rvec - \rvec', \partial) \rightarrow \delta (\rvec - \rvec') \Rmij (\partial)$. In reciprocal space, such that $\partial \rightarrow i \qvec$, we may then write the relationship between equilibrium and rigidity maps as
\begin{align}
\Qijm (\qvec)  = \Rmij ( -\qvec).
\end{align}

\noindent Our linear relationships then have the form:
\begin{align}
e_m(\qvec) = \Rmij(\qvec) \epij(\qvec), \\
\sij(\qvec) = \Qijm(\qvec) e_m(\qvec)
\end{align}

\section{Energy And Surface Terms}

In this section, we derive how the total energy of our system breaks down between surface and bulk terms, using the expression of our rigidity map.

\begin{subequations}
\begin{align}
E &= \frac{1}{2} \sum_m \int d^d \rvec \, e_m (\rvec)  e_m (\rvec)
\\
\text{with } e_m (\rvec) &=  \frac{b^m_i b^m_j}{|\bvec^m|} \left(1 + p^m_k \partial_k \right) \epij (\rvec)
\end{align}
\end{subequations}

\begin{subequations}
\begin{multline}
\implies E = \frac{1}{2} \sum_m \int d^d \rvec \, \frac{b^m_i b^m_j b^m_k b^m_l}{|\bvec^m|^2} \left[ (1 + p^m_\alpha \partial_\alpha) \epij (\rvec) \right] \times
\\
\left[(1 + p^m_\beta \partial_\beta) \epkl (\rvec) \right]
\end{multline}
\begin{multline}
\label{eqsi:derivation}
= \frac{1}{2} \sum_m  \frac{b^m_i b^m_j b^m_k b^m_l}{|\bvec^m|^2} \, \Bigg( \int d^d \rvec \, \bigg[ \epij(\rvec) (\pvec_m \cdot \nabla) \epkl (\rvec)
\\
+ \epkl(\rvec)  (\pvec_m \cdot \nabla) \epij (\rvec) \bigg] 
\\
+ \int d^d \rvec \, \left[\epij(\rvec) \epkl(\rvec) +  (\pvec_m \cdot \nabla) \epij(\rvec) \: (\pvec_m \cdot \nabla) \epkl(\rvec) \right] \Bigg)
\end{multline}
\end{subequations}

\noindent The second term of Eq.~(\ref{eqsi:derivation}) is a bulk term and can't be simplified any further. However the first term can nicely be expressed on the surface only using the divergence theorem:
\begin{subequations}
\begin{align}
E_s&=  \frac{1}{2} \sum_m  \frac{b^m_i b^m_j b^m_k b^m_l}{|\bvec^m|^2} \, \pvec_m \cdot \left( \int d^d \rvec \, \nabla (\epij (\rvec) \epkl (\rvec)) \right)
\\
&=  \frac{1}{2}\sum_m \int_{\textrm{surface}} d^{\dim-1} \rvec \left( \pvec_m \cdot \nvec\right) \frac{\bvecc^m_i \bvecc^m_j \bvecc^m_k \bvecc^m_l}
{|\bvec^m|^2} \strain_{ij} (\rvec) \strain_{kl} (\rvec).
\end{align}
\end{subequations}

\section{Winding Number}

As described in the main text, the task of relating the number of modes on edges of the systems to their bulk systems reduced mathematically to counting the numbers of zeroes in the complex plane. Here, we supply mathematical details and quantify the error associated with the long-wavelength approximation. This section implicitly uses a length scale so that we may treat physical quantities with units of length as pure numbers.

Physically, the points in the complex plane correspond to complex wavevectors. We draw a contour as shown in Fig.~\ref{fig:winding} of the main text which encloses all zeroes close to the origin (of order of the wavevector, which we term here $O(\epsilon)$) and avoids those far from the origin (of order 1). The latter are non-physical, since our theory only describes the long-wavelength limit. Due to the shape of the contour, zeroes in the upper half-plane, corresponding to modes on the left edge, are enclosed in a positive orientation, while those in the lower half-plane are enclosed in a negative one. Note also that the contour in question allows a branch cut to be defined stretching along the negative real axis, important for evaluating the phase of complex numbers.

Were we to retain all components of the contour, the Residue Theorem would ensure that we could exactly evaluate the number of zeroes enclosed by the contour (we choose a gauge in which no poles are present). However, in order to achieve bulk-boundary correspondence, we must identify the complex zeroes by only considering the bulk modes, which lie on the real axis. We thus neglect the curved components of the contour. We now show that this approximation nevertheless recovers the result given in the main text, up to a small, controlled error.

Without loss of generality, we may consider the phase added to our contour integral from each zero separately. As such, we wish to obtain the change in phase of a complex function of the form

\begin{align}
f(z) = z - z_0,
\end{align}

\noindent as $z$ winds along a contour $r e^{i\theta}$ with $\theta$ going  from 0 to either $\pi$ or $-\pi$. Here, we will choose $r$ to be much larger than long-wavelength zeroes ($O(\epsilon^1)$) but much smaller than the short-wavelength ones  ($O(\epsilon^0)$). 

Because the argument of the complex number is simply the imaginary part of its logarithm, we immediately obtain two results. First, the contribution of the straight sections of the contour, which we retain, is the expression given in the main text. Second, that the contribution from the curved parts which we neglect is

\begin{align}
2\,\textrm{Im} \log \left| \frac{r+z_0}{r-z_0}\right|.
\end{align}

Noting here that although the long-wavelength zeroes have real part $O(\epsilon^1)$ they have imaginary parts $O(\epsilon^2)$, we find that for $|r| \gg |z_0|$ the error is of order $O(\epsilon^2/r)$. In contrast, the error from the short-wavelength, spurious modes is of order $O(r)$, assuming that our original contour actually encloses all of the long-wavelength zero modes. However, if this condition is not met, the error in the calculation of the change in phase increases abruptly to $O(1)$. To minimize error, we should then select the bounds of our contour to minimize their size while ensuring that they capture the essential physics.
 In the main text, we make the natural choice for $r$ to lie intermediate between  the two regimes, $r = \epsilon^{1/2}$ in units in which unit cell size is of order one, generating error of $O(\epsilon^{1/2})$. More aggressive schemes, such as $r = \epsilon^{3/4},\epsilon^.99$ could more closely characterize the topological transition, but would require greater knowledge of the microscopic details or willingness to tolerate occasional large errors.

As discussed above, the error from the true determinant function may be obtained simply by summing over a few cases of these zeroes, and is thus of the same order. Thus, we have shown using complex analysis that in the long-wavelength limit the correspondence between the bulk structure and the imbalance of topological edge modes given in the main text is mathematically justified. We emphasize here that this is not because the bulk modes along the curved parts of the contour are negligible but because the choice of contour causes them to cancel out. Thus, bulk-boundary correspondence in the long-wavelength limit links together two boundaries at once, in contrast to the lattice theory.

\section{Soft Directions, Length Scales and choice of mathematical basis}

In the main text, we introduce a number of different bases for the wavevectors. First, we have a simple Cartesian basis $(\qh_x, \qh_y)$. Second, we present a rotated version thereof, in order to accommodate boundaries with varying orientations, in terms of inward-facing normal direction $\qh_n$ and the tangent direction $\qh_t$. Finally, we consider the two soft directions, along which there lie modes which satisfy the structural constraints to linear order, pointing along (non-orthogonal directions) $(\qh_+, \qh_-)$. We use the symbols $\alpha_\pm, \alpha_\pm (\ang)$ to relate these:
\begin{align}
\qh_\pm &= \frac{1}{\sqrt{1+\alpha^2_\pm}} \left( \alpha_\pm \qh_x + \qh_y \right) = \frac{1}{\sqrt{1+\alpha^2_+(\ang)}} \left( \alpha_+ (\ang) \qh_n + \qh_t \right).
\end{align}

\noindent As discussed in the main text, the bulk structure requires that zero modes occur at wavevectors satisfying the following constraint:
\begin{align}
q_x = \alpha_\pm q_y + i \bcoef \pm q_y^2.
\end{align}

\noindent We now wish to consider a boundary mode which has component $q_t$ along the tangent direction and thus takes the form  $q_n \qh_n + q_t \qh_t$. By requiring that this mode satisfies the above constraints to \emph{second} order, we obtain a result for the coefficients $\alpha_\pm(\ang)$ which define the soft directions:
\begin{align}
q_n &= \alpha_\pm (\ang) q_t + i \bcoef_\pm (\ang) q_t^2.
\end{align}

\noindent In the $\qh_n, \qh_t$ basis, we know the expression of $\qvec$. Let's assume that $\qvec$ is a zero of our map, therefore it has the following form:
\begin{align}
\qvec = q_t \left(\alpha_+ (\ang) \qh_n + \qh_t \right) + i \kappa_+ (\ang) q_t^2 \qh_n,
\label{eqsi:qvecnt}
\end{align}

\noindent We recognize that the leading-order contribution to this zero-energy edge mode is simply the bulk soft mode. Expressing it in this form explicitly, we may write:
\begin{subequations}
\begin{align}
\qvec &= \frac{q_t}{\qh_+ \cdot \qh_t} \qh_+ \, + \, i \kappa_+ (\ang) q_t^2 \qh_n,
\label{eqsi:qvect+} \\
\qvec &= q_+ \qh_+ \, + \,  i \kappa_+ (\ang) \left(\qh_+ \cdot \qh_t \right)^2 q_+^2 \qh_n,
\label{eqsi:qvec++}
\end{align}
\end{subequations}

\noindent where $q_+$ represents the value of the mode along the soft direction, while $q_t$ is the value of the mode along the transverse direction of the normal. Consider now such mode which, let's say, lies to first order along the $\qh_+$ direction. For it to be a zero mode, it must contain a second order correction along the $\qh_-$ direction that satisfies the zero-energy condition, which in this basis has a simplified expression:
\begin{multline}
\det \, \qvec = A''_{1,1} q_+ q_- + i (A''_{3,0} q_+^3 + A''_{2,1} q_+^2 \\
q_- + A''_{1,2} q_- q_+^2 + A''_{0,3} q_-^3) = 0.
\end{multline}

\noindent Based on a method relying on perturbation theory and using trigonometric identities, we're able to express the value of such zero mode as a function of the system's parameters only: 
\begin{subequations}
\begin{align}
\qvec &= \frac{q_t}{\sin (\theta_+ - \ang)} \qh_+ \, + \,
i q_t^2 \frac{A''_{3,0}}{A''_{1,1}}  \frac{\sin (\theta_+ - \theta_-)}{\sin^3 (\ang - \theta_+)} \qh_n
\label{eqsi:qvect+2}
\\
\qvec &= q_+ \qh_+ \, + \,
i q_+^2 \frac{A''_{3,0}}{A''_{1,1}}  \frac{\sin (\theta_+ - \theta_-)}{\sin (\ang - \theta_+)} \qh_n
\label{eqsi:qvec++2}
\end{align}
\end{subequations}

Finally, as we explain in the main, we find that some of the system's parameters are dependent of the size of the cell $|\lvec_1|$, particularly that the quantity $A''_{3,0} / (A''_{1,1} |\lvec_1|)$ is dimensionless. In addition to expressing the wavevector $q_t$ in terms of its wavelength $\lambda$ ($q_t = 2 \pi / \lambda$), we can then derive a expression for a zero mode as a function of the properties on the boundary and the associated decay length through the bulk $\zeta_+$:
\begin{align}
&\qvec = \frac{2 \pi}{\lambda \sin (\theta_+ - \ang)} \qh_+ +
i \left(\frac{2 \pi}{\lambda}\right)^2 \frac{A''_{3,0}}{A''_{1,1}}   \frac{\sin (\theta_+ - \theta_-)}{\sin^3 (\ang - \theta_+)} \qh_n
\label{eqsi:qvect+3}
\\
&\implies \frac{\zeta_+}{|\lvec_1|} =\left[\frac{\lambda}{2 \pi |\lvec_1|} \right]^2 \left[\frac{|\lvec_1| A''_{1,1}}{A''_{3,0}}\right] \frac{\sin^3 (\ang - \theta_+)}{\sin (\theta_+ - \theta_-)}
\label{eqsi:zeta}
\end{align}

\end{appendices}

\end{document}